\documentclass[12pt]{article}
\usepackage{epsf}
\title{Quark distributions in mesons in extended QCD\\
sum rule approach.}
\vspace{2cm}
\author{B.L.Ioffe\\
\\
Institute of Theoretical and Experimental Physics\\
B.Cheremushkinskaya 25, 117218, Moscow, Russia}
\date{}
\begin{document}

\maketitle

\def\Journal#1#2#3#4{{#1} {\bf #2}, #3 (#4)}

\def\NCA{\em Nuovo Cimento}
\def\NIM{\em Nucl. Instrum. Methods}
\def\NIMA{{\em Nucl. Instrum. Methods} A}
\def\NPB{{\em Nucl. Phys.} B}
\def\PLB{{\em Phys. Lett.}  B}
\def\PRL{\em Phys. Rev. Lett.}
\def\PRD{{\em Phys. Rev.} D}
\def\ZPC{{\em Z. Phys.} C}

\def\st{\scriptstyle}
\def\sst{\scriptscriptstyle}
\def\mco{\multicolumn}
\def\epp{\epsilon^{\prime}}
\def\vep{\varepsilon}
\def\ra{\rightarrow}
\def\ppg{\pi^+\pi^-\gamma}
\def\vp{{\bf p}}
\def\ko{K^0}
\def\kb{\bar{K^0}}
\def\al{\alpha}
\def\ab{\bar{\alpha}}
\def\be{\begin{equation}}
\def\ee{\end{equation}}
\def\bea{\begin{eqnarray}}
\def\eea{\end{eqnarray}}
\def\CPbar{\hbox{{\rm CP}\hskip-1.80em{/}}}

\def\la{\mathrel{\mathpalette\fun <}}
\def\ga{\mathrel{\mathpalette\fun >}}
\def\fun#1#2{\lower3.6pt\vbox{\baselineskip0pt\lineskip.9pt
\ialign{$\mathsurround=0pt#1\hfil##\hfil$\crcr#2\crcr\sim\crcr}}}

\vspace{5mm}

\begin{abstract}
The improved calculation method of quark distributions in hadrons in the
framework of QCD sum rules is presented.
The imaginary part of the virtual photon scattering amplitude on some
hadronic current is considered in the case, when initial and final
virtualities of the current $p^2_1$ and $p^2_2$  are different, $p^2_1\not=
p^2_2$. The double Borel transformation in $p^2_1, p^2_2$ is applied to the
sum rule, killing background non-diagonal transition terms, which
deteriorated the accuracy in the previous calculations. Valence quark
distributions in the pion were found in a good agreement with the data,
determined from the Drell-Yan process.

\end{abstract}

\vspace{5mm}
In my talk I present the recent results in determination of quark
distributions in mesons by using the improved QCD sum rule approach,
which were obtained by A.Oganesian and myself \cite{1}. I recall the idea of
the method \cite{2,3}. Consider the 4-point correlator of two
electromagnetic currents and two currents with quantum numbers of some
hadron  (for clarity the axial current, corresponding to charged pions is
considered):

\be
\Pi_{\mu\nu\lambda\rho}(p_1,p_2;q_1,q_2) = -\int
e^{ip_1x+iq_1y-ip_2z}d^4xd^4yd^4z\langle 0\mid T \left \{j_{5\lambda}(x)
j^{em}_{\mu}(y)j^{em}_{\nu}(0)j_{5\rho}(z)\right \}\mid 0 \rangle
\label{1}
\ee
\be
j_{5\lambda}=\bar{u}\gamma_5\gamma_{\lambda}d
\label{2}
\ee
where $q_1,q_2$ and $p_1,p_2$ are the photon and hadron current momenta,
correspondingly. Assume, that $q^2$ and $p^2$  are negative,
$q^2_1=q^2_2=q^2,~\mid q^2\mid \gg \mid p^2_1 \mid $, $\mid p^2_2 \mid$ and
$\mid p^2_1 \mid \sim \mid p^2_2 \mid \gg R^{-2}_c$, where $R_c$ is the
confinement radius. As was shown (\ref{3}), in this case the
imaginary part in $s$-channel of the correlator (\ref{1}) is dominated by
small distances in all channels.  Particularly, at $p_1=p_2$, $q_1=q_2$ the
closest to zero singularity of $Im~\Pi_{\mu\nu\lambda\rho}$  in the
$t$-channel is given by $(x=-q^2/2\nu,~\nu=pq)$

\be
t=-4\frac{x}{1-x}p^2,
\label{3}
\ee
and $t\sim \mid p^2 \mid \gg R^{-2}_c$ at nonsmall $x$, unlike the case of
$\Pi_{\mu\nu\lambda\rho}$ itself (the forward scattering amplitude), where
$t=0$ (for massless quarks)  and large distances in the $t$-channel are of
importance. This fact allows one to use operator product expansion (OPE) in
$1/p^2$ for calculation of $Im~\Pi_{\mu\nu\lambda\rho}$ in QCD, besides the
standard OPE in $1/q^2$. As follows from (\ref{3}) the approach does not work
at small $x$. It does not work also at large $x$, close to 1. Physically it
is evident, because this is the resonance region.

Phenomenologically $Im~\Pi_{\mu\nu\lambda\rho}$  is represented by
contributions of physical states using the double dispersion relation in
$p^2_1,~p^2_2$. Among various tensor structures of  $\Pi_{\mu\nu\lambda\rho}$
it is convenient to consider the structure
($P_{\mu}P_{\nu}P_{\lambda}P_{\rho}/\nu)\tilde{\Pi}(p^2_1,p^2_2,q^2,x)$,
where $P=(p_1+p_2)/2$, $x=-q^2/2(q_1P)$. The double dispersion
representation of $Im~\tilde{\Pi}$  has the form:

\be
Im~\tilde{\Pi}(p_1^2,p^2_2,x) = a(x) + \int\limits^{\infty}_0
\frac{\varphi(x,u)}{u-p^2_1} du +
\int\limits^{\infty}_0\frac{\varphi(x,u)}{u-p^2_2}+
\int\limits^{\infty}_0
du_1\int\limits^{\infty}_0
du_2\frac{\rho(x,u_1,u_2)}{(u_1-p^2_1)(u_2-p^2_2)}
\label{4}
\ee
(the  $q^2$ dependence is omitted). In the previous treatment of the
problem \cite{3} the dispersion representation (\ref{4}) was considered in
the limit $p^2_1=p^2_2=p^2$ and the single Borel transformation in $p^2$ was
performed. This results in appearance of nondesirable background terms,
which deteriorate  the accuracy of calculations and in the case pion even do
not allow one to find the quark distributions at all. The spectral functions
in (\ref{4}) were represented by contributions of the lowest state (pion)
and continuum

$$
\rho(u_1,u_2,x) =
f^2_{\pi}\cdot 2\pi F_2(x)\delta(u_1-m^2_{\pi})\delta(u_2-m^2_{\pi})
+\rho^0(x)\theta(u_1-s_0)\theta(u_2-s_0)$$
\be
\varphi(x,u) =  \varphi_1(x)\delta(u-m^2_{\pi}) +\varphi_2(x) \theta(u-s_0)
\label{5}
\ee
where $f_{\pi}=131~MeV$ and $F_2(x)$ is the pion structure function, $s_0$
is continuum threshold. $\varphi_1(x)$ corresponds to the contribution of
the nondiagonal transition $\gamma^*+\pi\to\pi^*+\gamma^*$ and is unknown.
It is not suppressed in comparison with the main term $\sim F_2(x)$ by the
single Borel transformation and must be accounted in the final sum rule.

The lowest order term of OPE corresponds to the box diagram of Fig.1. At
$p_1=p_2$ its contribution to $Im~\tilde{\Pi}$  is equal

\be
Im~ \tilde{\Pi}(p^2,x) = -\frac{3}{\pi}\frac{1}{p^2}x^2(1-x)
\label{6}
\ee
Using (\ref{6}), (\ref{8}) and (\ref{5})  and performing the single Borel
transformation in $p^2$  we get the sum rule ($M^2$ is the Borel parameter)

\be
\frac{3}{\pi}x^2(1-x)(1-e^{-s_0/M^2}) = 2\pi f^2_{\pi}x
u_{\pi}(x)\frac{1}{M^2} + \varphi_1(x),
\label{7}
\ee
where $u_{\pi}(x)$  is the distribution of valence $u$ quarks in the pion
(the pion mass is neglected). Looking at $M^2$ dependence  in (\ref{7})  it
becomes evident, that in this approach the attempt to separate the pion
contribution from the background by studying $M^2$  dependence (e.g.
differentiation over $1/M^2$) is useless -- up to small correction $\sim
e^{-s_0/M^2}$ the box diagram contributes to the background only.

Let us now improve the method  by considering the case $p^2\not=p^2_2$  and
performing the double Borel transformation in $p^2_1,p^2_2$. The double
Borel transformation kills the nondesirable first three terms in (\ref{4}).
Instead of (\ref{6})  we have now

\be
\tilde{\Pi}(p^2_1,p^2_2,x) = \frac{3}{\pi} x^2(1-x)
\int\limits^{\infty}_0du  \int\limits^{\infty}_0  du^{\prime}
\frac{\delta(u-u^{\prime})}{(u-p^2_1)(u^{\prime}-p^2_2)}
\label{8}
\ee
and after double Borel transformation with parameters $M^2_1$, $M^2_2$ the
sum rule arises

\be
u_{\pi}(x) = \frac{3}{2\pi^2}\frac{M^2}{f^2_{\pi}}x (1-x)(1-e^{-s_0/M^2}),
\label{9}
\ee
where it was put $M^2_1=M^2_2=2M^2$. The calculation of the pion decay
constant $f_{\pi}$, performed in \cite{4} in the same approximation leads to

\be
f^2_{\pi} = \frac{1}{4\pi^2} M^2(1-e^{-s_0/M^2})
\label{10}
\ee
The substitution of (\ref{10})  into (\ref{9})  gives

\be
u_{\pi}(x) = 6 x(1-x)
\label{11}
\ee
Therefore the necessary conditions

\be
\int\limits^1_0 u_{\pi}(x)dx = 1,~~~
\int\limits^1_0 x u_{\pi}(x)dx = 1/2
\label{12}
\ee
are fulfilled in this approximation.

Account now the perturbative and nonperturbative corrections. Restrict
ourselves to LO perturbative  corrections, proportional to $ln(Q^2/M^2)$ and
choose $Q^2=Q^2_0=2~GeV^2$  as a point at which we calculate the quark
distributions. The LO perturbative correction results in multiplication of
bare loop distribution (\ref{9}) by the factor

\be
r(x,Q^2)= \Biggl [1
+ \frac{\alpha_s(M^2) ln (Q^2/M^2)}{3 \pi} (1/x + 4 ln(1 - x) -
\frac{2(1 - 2x) ln x}{1 - x}) \Biggr ]
\label{13}
\ee

The higher order terms of OPE in $1/p^2$ starts from contribution of gluonic
condensate $\langle 0\mid G^n_{\mu \nu} G^n_{\mu \nu}\mid 0 \rangle $ of
dimension 4. The calcukation was performed in the Fock-Schwinger gauge
$x_{\mu}A^n_{\mu}(x)=0$  using the program of analytical calculation REDUCE.
Surprisingly, the sum of all diagrams, proportional to gluonic condensate
vanishes after double Borelization and gluonic condensate does not
contribute to the sum rule.

One-particle irreducible diagrams, (no loop diagrams) resulting in
appearance of $\delta(1-x)$ in $Im~\Pi$ and the diagrams, arising from their
QCD evolution are disregarded because the calculation method is inapplicable
at $x=1$. There are two vacuum expectation values of dimension 6: $\langle
g^3 f^{abc} G^a_{\mu\nu}G^b_{\nu\lambda}G^c_{\lambda\mu}\rangle$  and
$\langle \bar{\psi}\psi \rangle^2$ (for $\bar{\psi} 0\psi \cdot \bar{\psi}0\psi$
operators the factorization hypothesis was used). The calculation shows,
that the contribution of $G^3$ operators vanishes after double Borel
transformation and only $\langle \bar{\psi}\psi \rangle^2$ contribute at
$d=6$. The fonal result for valence $u$-quark distribution in $\pi^+$ with
account of LO perturbative corrections and OPE up to $d=6$  is given by

\be
x u_{\pi}(x) = \frac{3}{2\pi^2} \frac{M^2}{f^2_{\pi}}x^2(1-x)\cdot \Biggl
[ r (x,Q^2_0)
\cdot (1-e^{-s_0/M^2}) - \frac{4\pi\alpha_s(Q^2_0)\cdot 4\pi
\alpha_s(M^2)a^2}{(2\pi)^4\cdot 3^7\cdot 2^6 \cdot M^6}\cdot
\frac{\omega(x)}{x^3(1-x)^3}\Biggr ]
\label{14}
\ee
where $\omega(x)$  is the polynomial of 4-order in $x,a=-(2\pi)^2\langle
\bar{u}u\rangle$. The analysis of the sum rule (\ref{14}) shows, that it is
fulfilled at $0.15 < x < 0.7$; the power corrections are less than 30\% and
the continuum contribution is less than 25\%. Stability in $M^2$  at $0.4
<M^2<0.6~GeV^2$  is good. The final result for $x u_{\pi}(x)$ (at
$M^2=0.45~GeV^2$ and $s_0=0.8~GeV^2$) is shown in Fig.2. Fig.2  shows also
the curve of the valence $u$-quark distribution in the pion, found in
\cite{5} from the fit to the Drell-Yan procces data. The agreement is
satisfactory, especially bearing in mind, that nonaccounted NLO perturbative
correction would increase $u_{\pi}(x)$  at small $x$ and decrease at large
$x$. If we make the assumption that at $x \la 0.15 ~u_{\pi}(x)\sim
1/\sqrt{x}$ according to Regge behavior, and at $x \ga 0.7$ $u_{\pi}(x)\sim
(1-x)^2$, then the moments can be found

\be
{\cal{M}}_0 = \int\limits^{1}_{0} u_{\pi}(x) dx \approx 0.84~~~~~
{\cal{M}}_1 =  \int\limits^{1}_{0} xu_{\pi}(x) dx \approx 0.21
\label{15}
\ee

This work was supported in part by RFBR grant 97-02-16131.

\newpage

\begin{figure}
\epsfxsize=5cm
\epsfbox{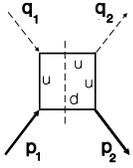}
\caption{Diagrams, corresponding to the unit operator contribution. Dashed 
lines with arrows correspond to the photon, thick solid - to hadron current}
\end{figure}

\begin{figure}
\epsfxsize=5cm
\epsfbox{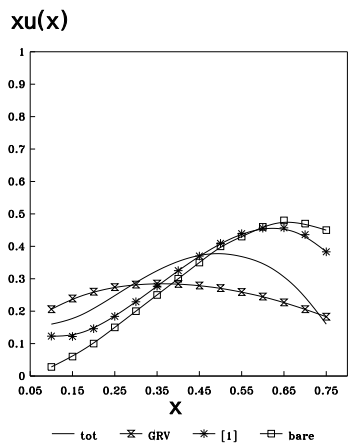}
\caption{Quark distribution function in pion, noted "total". For comparison
 fit from [5], noted "GRV", is shown. Also bare loop ("bare") and 
bare loop with nonperturbative corrections (noted "1"), are shown}
\end{figure}

\end{document}